# Optical loss compensation in a bulk left-handed metamaterial by the gain in quantum dots

Zheng-Gao Dong [a]

*Physics Department, Southeast University, Nanjing 211189, China*

Hui Liu,[b] Tao Li, Zhi-Hong Zhu, Shu-Ming Wang, Jing-Xiao Cao, and Shi-Ning Zhu

*National Laboratory of Solid State Microstructures, Nanjing University, Nanjing 210093, China*

X. Zhang

*25130 Etcheverry Hall, Nanoscale Science and Engineering Center, University of California, Berkeley, California 94720-1740, USA*

A bulk left-handed metamaterial with fishnet structure is investigated to show the optical loss compensation via surface plasmon amplification, with the assistance of a Gaussian gain in PbS quantum dots. The optical resonance enhancement around 200 THz is confirmed by the retrieval method. By exploring the dependence of propagation loss on the gain coefficient and metamaterial thickness, we verify numerically that the left-handed response can endure a large propagation thickness with ultralow and stable loss under a certain gain coefficient.

[a] Electronic address: zgdong@seu.edu.cn
[b] Electronic address: liuhui@nju.edu.cn; URL: http://dsl.nju.edu.cn/ dslweb/images/ plasmonics-MPP.htm

Three-dimensional (3D) optical metamaterials with artificial magnetic response will make the intriguing ideas, such as super imaging and cloaking, closer to actual truths.[1,2] The method to manufacture 3D optical metamaterials using a layer-by-layer technique, including fabrication processes of planarization, alignment, and stacking, has presented a breakthrough for building 3D optical metamaterials,[3] even for so-called stereometamaterial with flexible twists between layers.[4] As were adopted in literatures, left-handed metamaterial with metallic fishnet pattern is one of the most popular artificial structures for the studies about negative refraction, cloaking, and magnetic plasmon polaritions.[5-8] In contrast to a microwave counterpart, the fishnet structure shows striking advantage for the processing purpose of optical 3D metamaterial, what required is simply an array of perforated holes on metallic layers. Recently, negative index of refraction has been demonstrated experimentally in optical frequencies by constructing a bulk metamaterial with 21 layers of alternating $Ag/MgF_2$ fishnet structure.[9] However, metallic metamaterials in various shapes,[10-12] including the fishnet one, generally encounter heavy losses in propagating electromagnetic energy, particularly at optical frequencies.[13-16] Consequently, the issue of serious optical loss, or propagation deterioration, needs to be addressed in a practical bulk metamaterial with a large number of stacking layers.

To compensate the propagation loss and thus pave the way toward bulk optical metamaterials, resorting to the active medium is a readily available way. In contrast to the general amplification process by a gain medium,[17,18] surface plasmon amplification by stimulated emission of radiation, shortened as spaser,[19,20] has the brilliance to achieve potentially orders of magnitude enhancement around resonant narrowband with a moderate gain coefficient requirement. Recently, a lasing spaser was proposed to create a confined version of coherent source of electromagnetic radiation fuelled by plasmonic oscillations.[21-23] Obviously, surface plasmon amplification provides a potential way to overcome the propagation energy loss of electromagnetic waves in bulk metamaterials, since such artificial materials are generally metallic and of resonance. As a matter of fact, although there are increasing works about gain-assisted electromagnetic properties in literatures, little was concerned on the *bulk* left-handed propagation with loss compensated by an optically active system. In this Letter, we investigate numerically the resonance enhancement at optical frequencies in an active bulk fishnet metamaterial by amplifying the surface plasmon. By taking into account the realistic metal and active materials, the loss-compensated left-handed transmission can be achieved and is sustainable for sufficient propagation layers at a particular gain level.

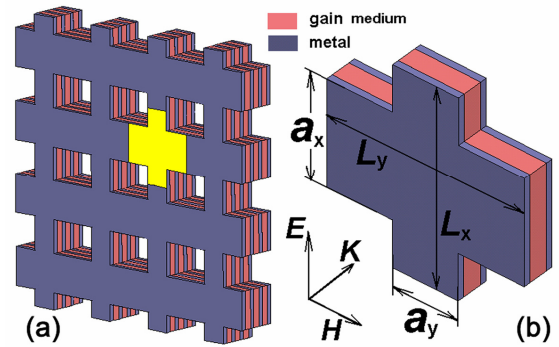

FIG. 1. (Color online) The schematic illustration of the bulk fishnet metamaterial with active layers as the spacer. (a) Bulk fishnet metamaterial with 4-layer metallic fishnet structure, the yellow area indicates an *xy*-face of the selected unit cell. (b) The unit symbols for the fishnet metamaterial.

Figure 1(a) shows schematically the bulk fishnet metamaterial with four metallic layers, where geometric parameters depicted in Fig. 1(b) are as follows: $a_x = 496\,\text{nm}$, $a_y = 310\,\text{nm}$, and $L_x = L_y = 930\,\text{nm}$. The 62-nm-thick metal in fishnet pattern is silver with the Drude dispersion ($\omega_p = 1.37 \times 10^{16}\,\text{s}^{-1}$ and $\gamma = 8.5 \times 10^{13}\,\text{s}^{-1}$, see Refs. 24 and 25). In addition, the spacer layer is Polymethyl methacrylate (PMMA, optical index of refraction $n=1.49$) with a layer thickness of 93 nm (i.e., the gap between the adjacent metallic layers). The gain medium imbedded in the PMMA spacer is PbS semiconductor quantum dots, characterized by the gain coefficient $\alpha = (2\pi/\lambda)\text{Im}(\sqrt{\varepsilon' + i\varepsilon''})$, where $\varepsilon'$ and $\varepsilon''$ are the real and imaginary parts of the electric permittivity of the active system, respectively.[21] According to the emission property of PbS quantum dots, its gain coefficient is described in this Letter by a Gaussian distribution, with the maximum gain $\alpha_0$ at 1500 nm wavelength and a full width at half maximum (FWHM) 150 nm. As for the absolute value of the maximum gain $\alpha_0$ at 1500 nm, it is tunable and experimentally depends on the quantum dots density, pumping power, sample temperature, and so on.[26] In our numerical simulations based on the full-wave finite element method, the polarized incident wave with electric field component in the $x$ direction and magnetic field component in the $y$ direction is satisfied by applying the perfect electric and magnetic boundary conditions, respectively.[24,27] For clarity, the unit cell in Fig. 1(b) is referred to as a 2-layer fishnet structure, which is the primary functional stack of a fishnet metamaterial since no monolayer metallic fishnet was expected to exhibit a left-handed behavior. On the analogy of this, the illustration in Fig. 1(a) is a 4-layer fishnet metamaterial.

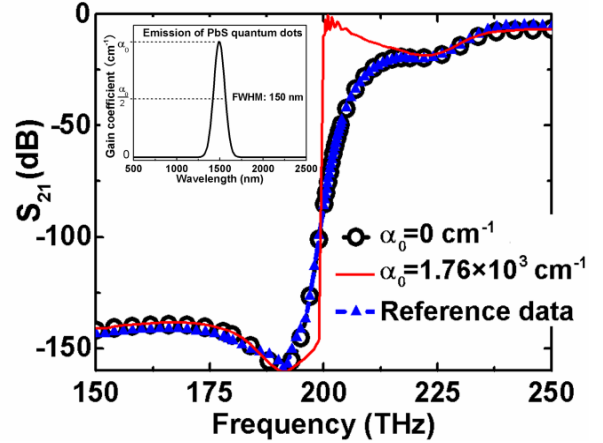

FIG. 2. (Color online) Transmission spectra of the 10-layer fishnet metamaterial with and without gain. The reference data are obtained with the same $\alpha_0 = 1.76 \times 10^3\,\text{cm}^{-1}$, but at an assumed emission wavelength 1250 nm for the active quantum dots. The inset shows the gain regime of PbS quantum dots.

Figure 2 shows the comparison results, with and without gain, of a 10-layer fishnet metamaterial, approximately the experimental stack numbers in Ref. 9. It is obvious that there is a resonance transmission near 200 THz by introducing the active medium of PbS quantum dots. On the one hand, the large amplification around 200 THz is not resulted directly from the emission of the gain medium. This can be confirmed by the reference data (solid triangular line in Fig. 2), which is the result simulated with the same maximum gain coefficient $\alpha_0 = 1.76 \times 10^3\,\text{cm}^{-1}$, but at an assumed emission wavelength of 1250 nm for PbS quantum dots. We can find from the reference line that such a gain level itself is not enough to reach an obvious enhancement as compared with the zero-gain transmission spectrum (open circular line in Fig. 2). On the other hand, the reference line in Fig. 2 also implies that the surface plasmon resonance from the metallic fishnet structure should play an important role in the sufficient transmission enhancement at a small gain level, as shown by the red curve in Fig. 2. Frankly speaking, it is far from enough to simply attribute the resonance enhancement to an effect of loss compensation by the gain.[23] Instead, a full understanding of this combined system should take into account the strong coupling between the quantum-mechanical (active quantum dots) and electrodynamic (resonant metamaterial) systems, which is hard to describe here. Nevertheless, a further investigation can be performed on the effective property of the gain-assisted fishnet metamaterial.

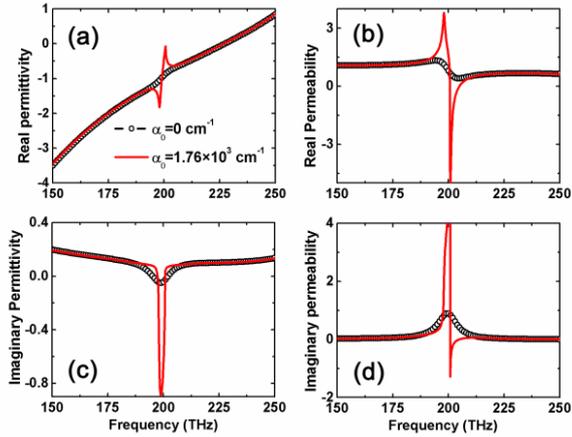

FIG. 3. (Color online) Retrieved constitutive parameters of the proposed fishnet metamaterial.

To confirm the amplified optical resonance and its left-handed characteristic in the proposed bulk fishnet metamaterial, the generally used retrieval procedure is practical since the fishnet periodicity is about $\lambda/10$ in the propagation direction. Without gain medium assisted, the magnetic resonance around 200 THz is very weak such that the real permeability loses its negative values [Fig. 3(b)]. In contrast, an amplified optical resonance with negative real permeability emerges under $\alpha_0 = 1.76 \times 10^3 \, cm^{-1}$. Additionally, Figs. 3(c) and 3(d) show the dispersions of the imaginary permittivity and permeability, respectively, which are consistent with the sign strategy for constitutive parameters in an active system.[18] Besides the simultaneously negative permittivity and permeability, it has been confirmed in our simulations that the resonant mode near 200 THz (not shown) has the same current distribution with a microwave version of the left-handed fishnet metamateiral.[6]

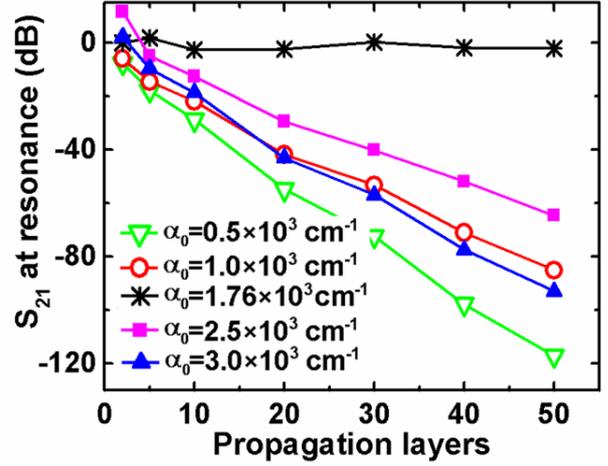

FIG. 4. (Color online) Dependence of the left-handed transmission on propagation layers and gain coefficient.

As has been demonstrated in Figs. 2 and 3, optical loss compensation of the left-handed propagation can be achieved in the 10-layer bulk fishnet metamaterial. However, whether or not the compensation effect is sustainable when the propagation thickness increases? What is the gain dependence of the bulk propagation of loss-compensated left-handed response? The simulation results shown in Fig. 4 can provide quantitative answers, from which it is interesting to find that, only for a certain gain coefficient level, a constant compensation effect, i.e., a stable transmission independence of the propagation layers, can be achieved even if the propagation thickness is increased up to 50-layer fishnet. On the contrary, for larger as well as smaller gain levels, the transmission deterioration at resonance would get more and more severe with the propagation thickness, indicating that the elongation of the surface plasmon propagation should not be overcompensated by the gain. However, it should be emphasized that the transmission reduction above the certain critical value of gain, though in accordance with literatures,[21,28] should be attributed to an unrealistic consequence of the time-independent solution.[29,30] In fact, physically there is no way to create and maintain such values of gain coefficient in a steady state. By solving time-dependent equations, this counterintuitive discrepancy can be explained satisfactorily.[29,30]

In summary, an optically active left-handed metamaterial in bulk fishnet stacks is investigated numerically by taking into account the realistic active medium of PbS quantum dots with Gaussian gain profile. It is indicated that tremendous resonant loss in the optical fishnet left-handed metamaterial, which could prevent bulk propagation of electromagnetic energy, can be reduced substantially under a moderate gain level. Additionally, stability of transmission magnitude can be established regardless of the propagation thickness. The results should be helpful for the experimental exploration of gain-assisted bulk metamaterials in the future.

This work was supported by the National Natural Science Foundation of China (Nos. 10534020, 10604029, 10704036, 10874081, and 10904012).